# Analog Computing revisited: A fully analog and minimalistic Damage Detector for Ultrasonic Testing enabling Material-Integrated Structural Health Monitoring

Stefan Bosse[1]
[1]University of Koblenz, Practical Computer Science, Koblenz, Germany

**Abstract**. Ultrasonic Testing (UT) is commonly used to detect damage in structures, e.g., metal plates. A sensor acquires Ultrasonic waves, e.g., by using PZT transducers. The time-resolved sensor signal must be processed with analog electronics, e.g., amplified and filtered. Commonly a digitalization follows using an Analog-to-Digital converter, finally processing the digital sensor signal, applying digital signal processing, feature extraction, and Machine Learning by using powerful microprocessor systems. The disadvantages of digital processing systems are their high number of transistors (microchip area), energy consumption, state-dependent processing and therefore sensitivity to energy supply interruption. Beyond silicon electronics, printed organic electronics gains interest. But printed electronics is still limited to low transistor and electronic component counts (typically 100). We will investigate and demonstrate a fully analog signal processing and feature extraction system consisting of an analog Hilbert transform deriving the signal envelope, simple analog arithmetic calculations for feature extraction, and finally damage classification and regression using an analog Artificial Neural Network. We expect a full damage detection system with less than 100 transistors. We will test our damage detection system with PZT transducer signals from Steel plates with circular defects. The focus of this work is the analog computation of the signal envelope (using all-pass filter networks for approximation of the Hilbert transform) and the analog feature extraction as well as the prediction of damage, forming an analog computer which can perform in-sensor computation, computing without a digital computer.

## 1. Introduction

Damage diagnostics relies on the processing and evaluation of sensor signals, e.g., widely deployed time-resolved Ultrasonic signals, with an example shown in Fig. 1. Detection of damage assumes that a damage changes the sensor signals in such way that damage-related features are added to the signal. There is commonly a baseline signal without any damage feature with a fundamental information content $I_0$ characterizing the test specimen and measuring set-up (e.g., signal frequency), compared with a signal containing an extended information content $I_0$+$I_d$. The difference signal will provide temporal information about the information density and the desired $I_d$. The significant part of the signal related to the identification of damage can be narrowed to a specific time range within the entire signal [1].

Ultrasonic testing uses in its minimal configuration a pitch (sender) and a catch (receiver) transducer with at least on scan path, which can be extended to more than two transducers and multiple scan paths [2].

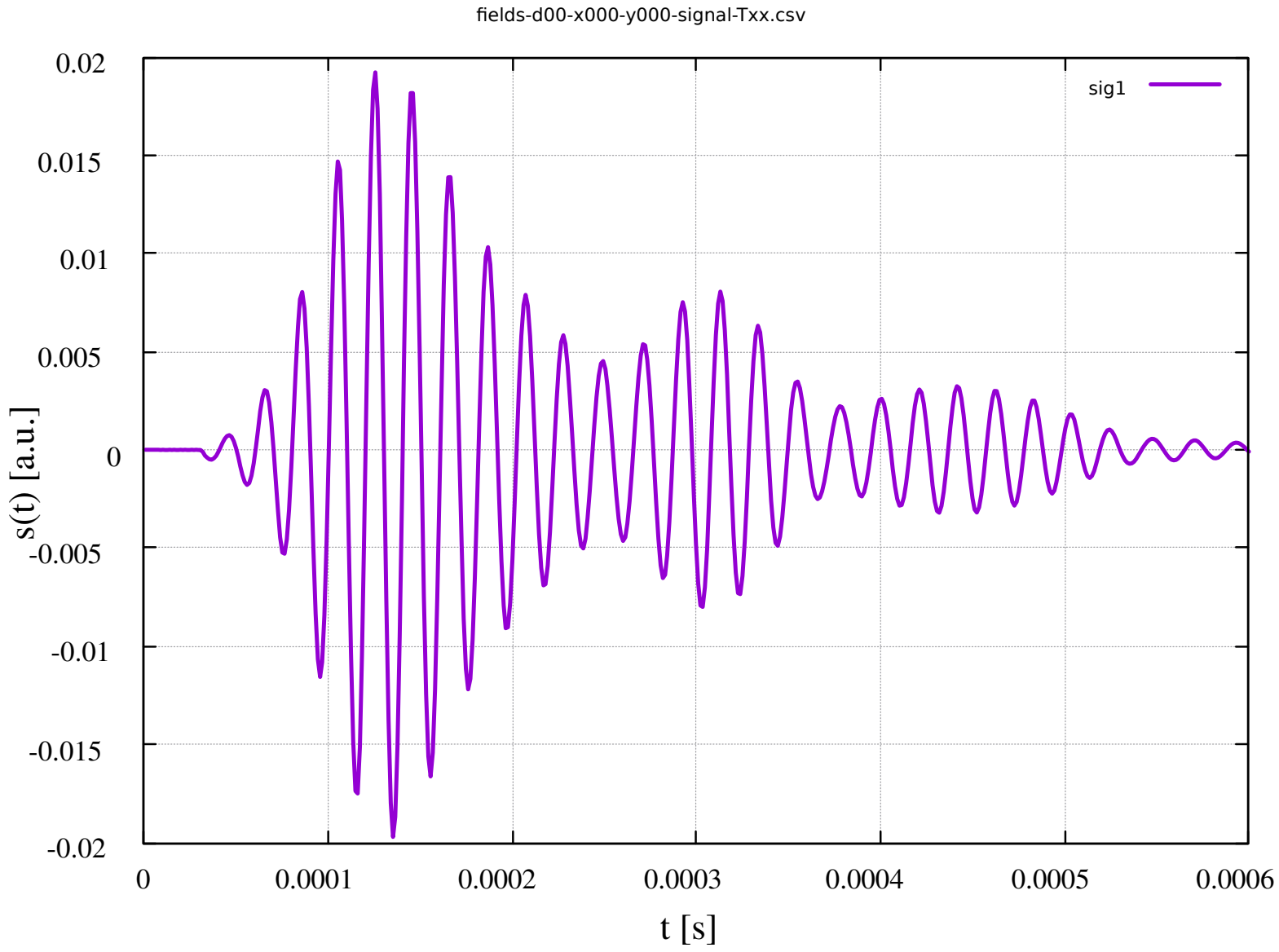

Fig. 1. A typical Ultrasonic catch signal captured with a piezo transducer. The signal was generated by a gaussian-window sine pitch signal ($\omega_0$=50kHz, $n$=10 cycles).

Sensor signals are mostly inherently analog, but commonly processed by digital computers, requiring an analog-digital conversion.

Contrary to the dominance of digital technology, there is currently a return to analog circuit principles for specific applications. Analog systems can provide advantages in terms of resource requirements and explainability. Analog circuits are often used for signal amplification and signal conditioning only, but are to be understood and used here generally as analog calculators for functional models. Functional calculations can be static without a state (e.g., linear algebra), or dynamic with a state (e.g., signal transformation in the time-domain, like frequency filters).

Analog circuits can perform a series of functional calculations approximated with acceptable loss of accuracy with a significantly reduced number of components, mainly transistors, and therefore generally have a reduced energy and area requirement. Functional calculations are an integral part of sensor signal processing, e.g., sensor calibration and fusion. Analog circuits can also contribute to improved explainability and validability compared to (software-based) digital circuits, since analog calculation circuits can represent an image of mathematical models (dualism and bijectivity), and in particular can have increased stability to disturbances in the energy supply (essentially in sensor nodes supplied autonomously via local energy generation).

The mapping of mathematical model functions to analog circuits can be done systematically by means of elementary cells, where the classical operational amplifier (OpAmp) is a long-used basic block with which almost all functional mappings and calculations f(x): x $\rightarrow$ y can be implemented systematically and analytically, as was shown and demonstrated by Ulmann [3], including multiplication, division (by inversion), logarithmizing and square root (by inversion), and solving algebraic problems. The OpAmp can therefore be seen as a powerful and universal elementary cell and functional systems can be synthesized directly with little algorithmic effort (functional composition).

Analog circuits are of high relevance for the establishment of organic electronics, an emerging key technology for flexible and material-integrated sensing systems. Organic sensors and transistors are ideally suited for material integration (with a focus on stretchability and bendability as Sekitani et al. described in [4] for artificial skin, and Bornemann et al. in [5] for structural monitoring) and body-related sensors for health care and medicine, such as Demuru et al. in [6] demonstrated. Not only do they offer high sensitivities, but they can also be printed, which enables rapid adaptation in the design process, or screen-printed over a large area [7].

This work is a proof-of-concept study that should investigate and answer the questions if it is possible to perform sensor signal processing, feature extraction, and damage prediction using only discrete analog electronics, typically with less than 100 transistors. The future goal, although not addressed in this work directly, is the implementation of functional sensor signal processing with printed organic electronic circuits, i.e., bringing analog computation tightly coupled to sensors. We will show analog electronic circuit solutions for a variety of mathematical computations required for feature extraction of Ultrasonic sensor signals, finally processed by an analog artificial neural network used for damage prediction. Detailed electronic component values of all circuits are summarized in Appendix A.

## 2. Hilbert Transform and Analytical Signal

There are different methods to extract relevant features from time-dependent signals. Relevant features are parts of the signal that carry the relevant information, i.e., here the damage features. Besides time-to-frequency domain transformations, i.e., Fourier or Wavelet transforms, the signal hull (envelope) is an important intermediate signal transformation, as shown in Fig. 2. The signal hull can be obtained from the analytical signal, which can be calculated by using the Hilbert transform.

The analytical signal $x_{\mathrm{a}}(t)$ of a real-valued sensor signal $x(t)$ is given by the superposition of the original signal and its complex-valued Hilbert transform:

$$x_a(t) = x(t) + i\tilde{x}(t),\\ \tilde{x}(t) = H(x(t)) \tag{1}$$

The magnitude of the analytical signal $x_{\ae}$ gives the envelope or hull of a signal:

$$E(x,t) = |x_a(t)| = \sqrt{x^2 + \tilde{x}^2} \tag{2}$$

A rough but for this work sufficient approximation of the magnitude calculation without squaring can be achieved with:

$$E(x,t) = |x_a(t)| \approx |x| + |\tilde{x}| \tag{3}$$

This simplification is important for the following analog electronic circuits computing the Hilbert transform and the magnitude, i.e., the envelope signal, avoiding analog multipliers, which are the most challenging circuits if accuracy and stability matters. The simplified magnitude approximation is characterized by a loss of smoothness showing some ripple.

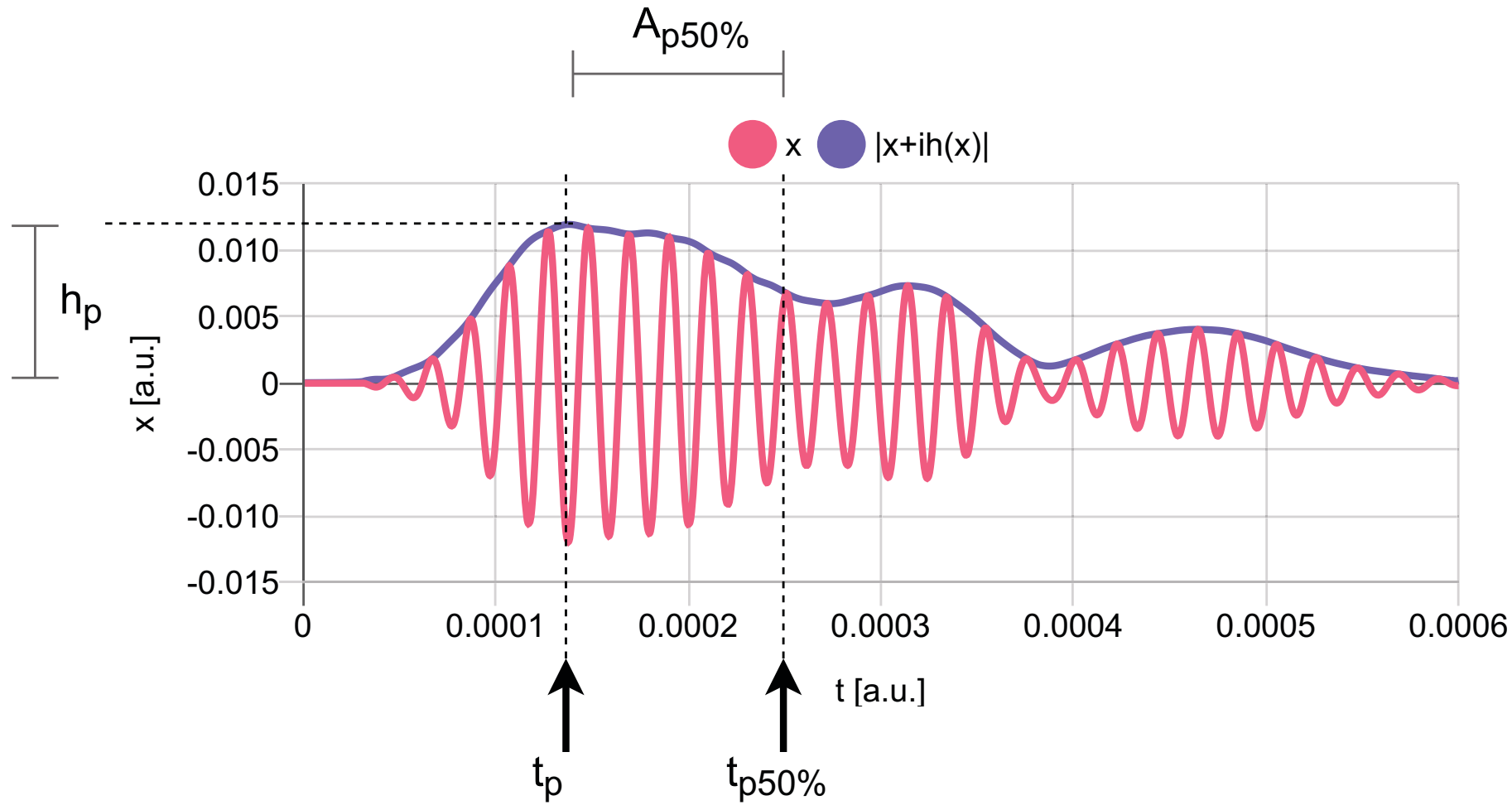

Fig. 2. Example of the envelope (hull) of an Ultrasonic signal computed by using the Hilbert transformation. Relevant features of the hull, mainly of the main peak, are shown.

The Hilbert transform performs basically a phase shift of the original real signal by 90° independent of the frequency of the signal (or its frequency components). This signal phase shift can be easily computed by an all-pass frequency phase shifter network, as rigorously investigated by Hutchins [8] and practically designed by Bode [9], as illustrated in Fig. 3.

The all-pass filters (first order) are formed from a cascade of one-pole one-zero all-pass filters, each having the following transfer function:

$$h(s) = \frac{s+p}{s-p} \tag{4}$$

where $p$ represents the location of the filter pole.

The number of all-pass shifters in each branch determines the approximation accuracy of the phase shift between the input and shifted signal. The design of the all-pass shifter networks A and B is constrained by a maximal phase error ε within a desired frequency range [$f_0$,$f_1$]. Hutchins [8] showed the advantage of second order all-pass filters (each with two poles), but second-order filters are practically not suitable for the analog processor with discrete transistor circuits as introduced in the next Section, and therefore we are still using first order filters.

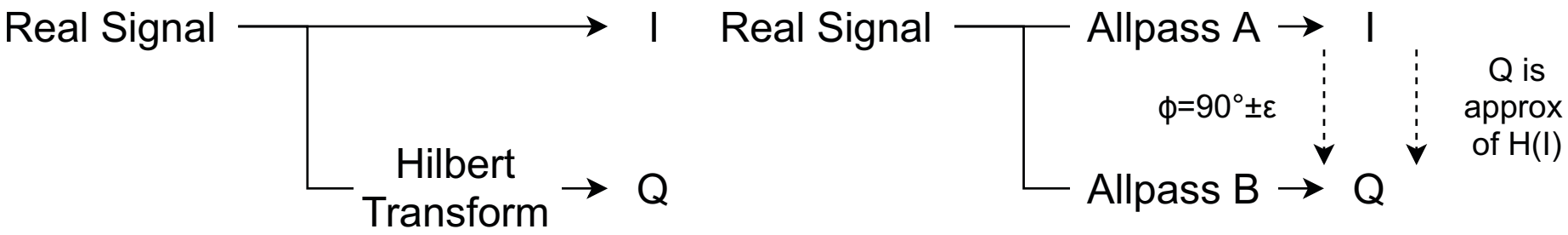

Fig. 3. Computing the analytical signal by using an all-pass frequency phase shifter as an approximation for the Hilbert transform Q≈H(I).

The accurate computation of the pole frequencies of the all-pass filters is crucial. Based on the work of Hutchins [8] we start by defining a relevant frequency range, e.g., for Ultrasonic signals $f_r$=[10 kHz, 300 kHz], and the fixed number of filter order $n$, e.g., $n$=4. This defines a frequency range bandwidth ratio $B$=$f_1$/$f_0$, which defines the spread of pole frequencies for path A and B filters.

We can then define (without proofing):

$$\begin{gathered} k = \sqrt{1 - \frac{1}{B^2}}, L = \frac{1}{2}\frac{1-\sqrt{k}}{1+\sqrt{k}} \\ P_0 = L + 2L^5 + 15L^9 \\ P = e^{\frac{\pi^2}{\ln{(P_0)}}} \end{gathered} \tag{5}$$

The maximal phase shift error over the desired frequency range can be estimated with (according to [8]):

$$\epsilon = \frac{720}{\pi} P^n, \tag{6}$$

assuming $P < 1$.

A phase shifter network of order $n$ requires $2n$ first order all-pass filters, with different pole frequencies divided into path $A$ and $B$. The pole frequencies $p$ can be computed by:

$$\begin{aligned}
\phi_{A,i} &= \frac{45}{n}(4i-3) \\
\phi_{B,i} &= \frac{45}{n}(4i-1) \\
P_{26} &= P^2 - P^6 \\
\theta_i &= \tan^{-1}\left(\frac{P_{26}\sin(4\phi_i)}{1+P_{26}\cos(4\phi_i)}\right) \\
p_i &= \sqrt{B}\tan(\phi_i - \theta_i),
\end{aligned} \tag{7}$$

for $i$=1,2,.., $n$, and $\phi_A$ for path $A$ and $\phi_B$ for path $B$, respectively (angles in degree). Example calculations for a bandwidth [10 kHZ, 300 kHz] are shown in Tab. 1.

| **i** | $p_A$ [Hz] | $p_B$ [Hz] |
|---|---|---|
| 1 | 3037 | 10236 |
| 2 | 21231 | 40176 |
| 3 | 74670 | 141301 |
| 4 | 293095 | 987520 |

Tab. 1. Pole frequencies for $n$=4 all-pass filter network providing 90° phase shift between branch A and B. .

## 3. Signal Processing Architecture

The goal of this work is the calculation of the envelope (hull) of a time-dependent signal, here an Ultrasonic signal used for damage diagnostics. Typical and relevant Ultrasonic signal features are related to the main peak (see also Fig. 2):

1. The (relative) time of the peak of the maximum main peak $t_p$;
2. The height of the maximum main peak $h_p$;
3. The width of the main peak at a specific fraction $k$ of the peak maximum height (e.g., $k$=0.5 as the full width at half maximum), here expressed as a (relative) time $t_{\mathrm{p\%}}$;

4. The area of the main peak at a specific fraction $k$ of the peak maximum height, $A_{\mathrm{p\%}}$.

The mathematical and digital calculation of these feature is rather trivial and straight forward. Calculating these features with an analog circuit is tricky and a challenge. The principle analog processing architecture is shown in Fig. 4.

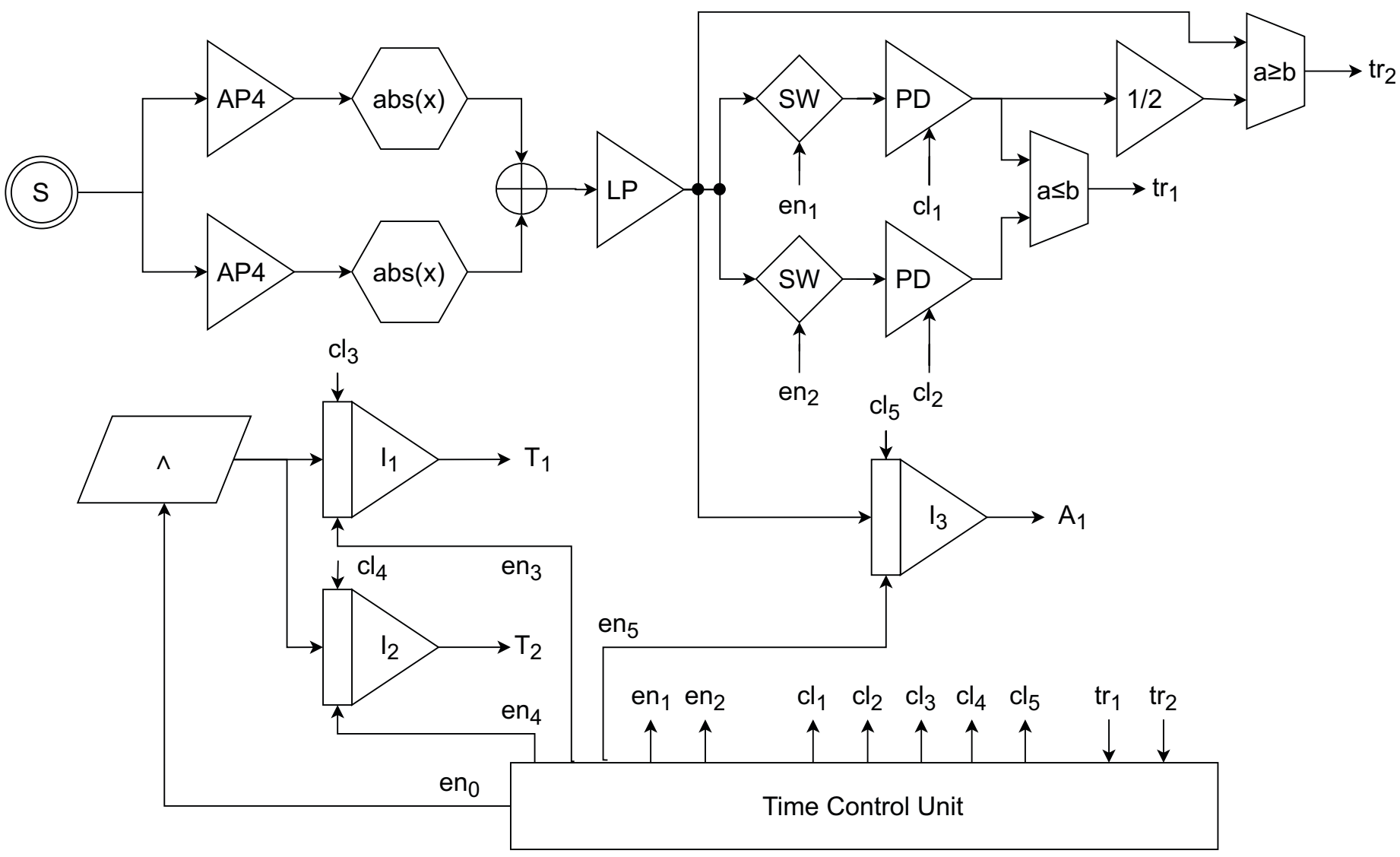

Fig. 4. The functional block diagram of the entire analog signal processing system (without damage classificator) S: Time-dependent input signal, AP: All-pass phase shifter filter, abs: Full wave rectifier, LP: Low pass filter, SW: Analog switch, PD: Peak detector, I: Integrator, Λ: Ramp generator.

The measured signal is passed to the all-pass shifter network AP (technically introduced in Sec. 4), merging both paths with an adder circuit. As already discussed, the envelope signal is noisy and shows ripple, which is reduced by a low-pass frequency filter LP. Via two controlled analog switches peak detectors will store the maximum height of the main peak (which is here the highest peak). In contrast to digital systems, we cannot play backward, therefore we need to adjacent signals, i.e., two measurements. In the first run (upper peak detector path), the peak maximum is stored. In the second run, the lower path of the peak detector is activated. Two comparators will deliver signals to the central timer control unit (TMU), controlling the integrators $I_1$ and $I_2$, recording the time points $t_p$ and $t_{p\%}$ by integrating a ramp generator signal (output signals $T_1$ and $T_2$, respectively). The third integrator $I_3$ delivers the area of the main peak starting from the peak to a fractional height of the peak (typically in the range of 20%-50%, output signal $A_1$). The output from the peak detector and the three integrators are used as analog signal features for the damage diagnostic system, mainly consisting of an analog artificial neural network, as discussed in Sec. 5.

## 4. Analog Signal Processing System: Feature Extraction

The central part of the analog feature extractor is the all-pass filter network. A minimal discrete transistor implementation of a fourth order ($n$=4) network is shown in Fig. 5. Each filter is build with a single transistor amplifier (unity gain), Q11-Q14 for path A, Q21-Q24 for path B, respectively. The filter frequencies are determined by the R1x, C1x, R2x, and C2x resistor-capacitor pairs. Because the resistors R1x and R2x provide a bias current for the following amplifier stage, these resistor values must be set to an equal specific value setting the static DC operation point of the amplifier. Therefore, the pole frequency is given only by the capacitor by using the following formula:

$$C_i = \frac{1}{2\pi R f_i}, f_i = p_i \tag{8}$$

As discussed in Sec. 2, second order filters (combination of two poles) are superior over first order filters (see [8]). But with single transistor amplifiers as used in this work it is not possible to decouple two RC networks from each other, and two RC networks would always interfere with each other, introducing significant phase errors. The single-order filter design still produces sufficient phase accuracy, as shown in Fig. 6, with a phase error below 1.5° within the specified bandwidth.

As outlined in the general analog processing architecture overview, we need some more mathematical function to calculate relevant signal features from the envelope of the signal. Most circuits base on difference amplifier stages and minimal OpAmp circuits. Fig. 7 shows a simple implementation of the absolute value function using an active full-wave rectifier. It consists of two minimal OpAmp circuits coupled via diodes. The OpAmp circuit linearize the non-linear diode voltage-current behavior. Each OpAmp stage consists of an input differential amplifier and an output current and voltage amplifier.

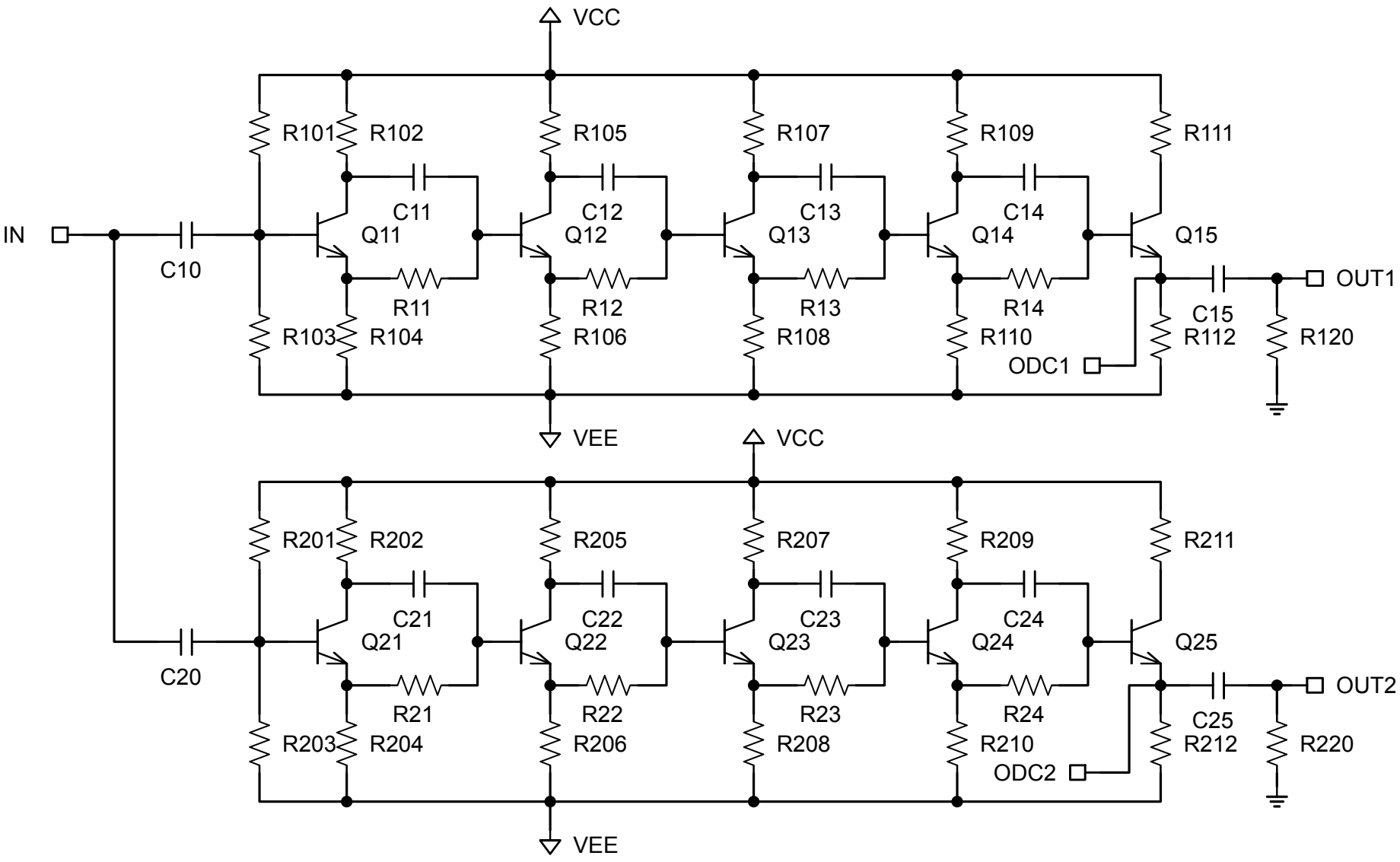

Fig. 5. A discrete all-pass fourth order phase shifter network (originally proposed by Bode [9]) implementing the function $y$=$H(x)$. The input signal IN is split into two signals OUT1 and OUT2 with unity gain and a phase shift between the two signals of 90°.

Fig. 8 shows the implementation of the maximum function using a modified version of a rectifier with a capacitive energy storage. To avoid leakage (decrease of the stored peak value with time) we used JFET transistors in the difference amplifier stage.

Fig. 9 shows the implementation of an adder circuit with additional low-pass frequency filters, again using a minimal OpAmp circuit.

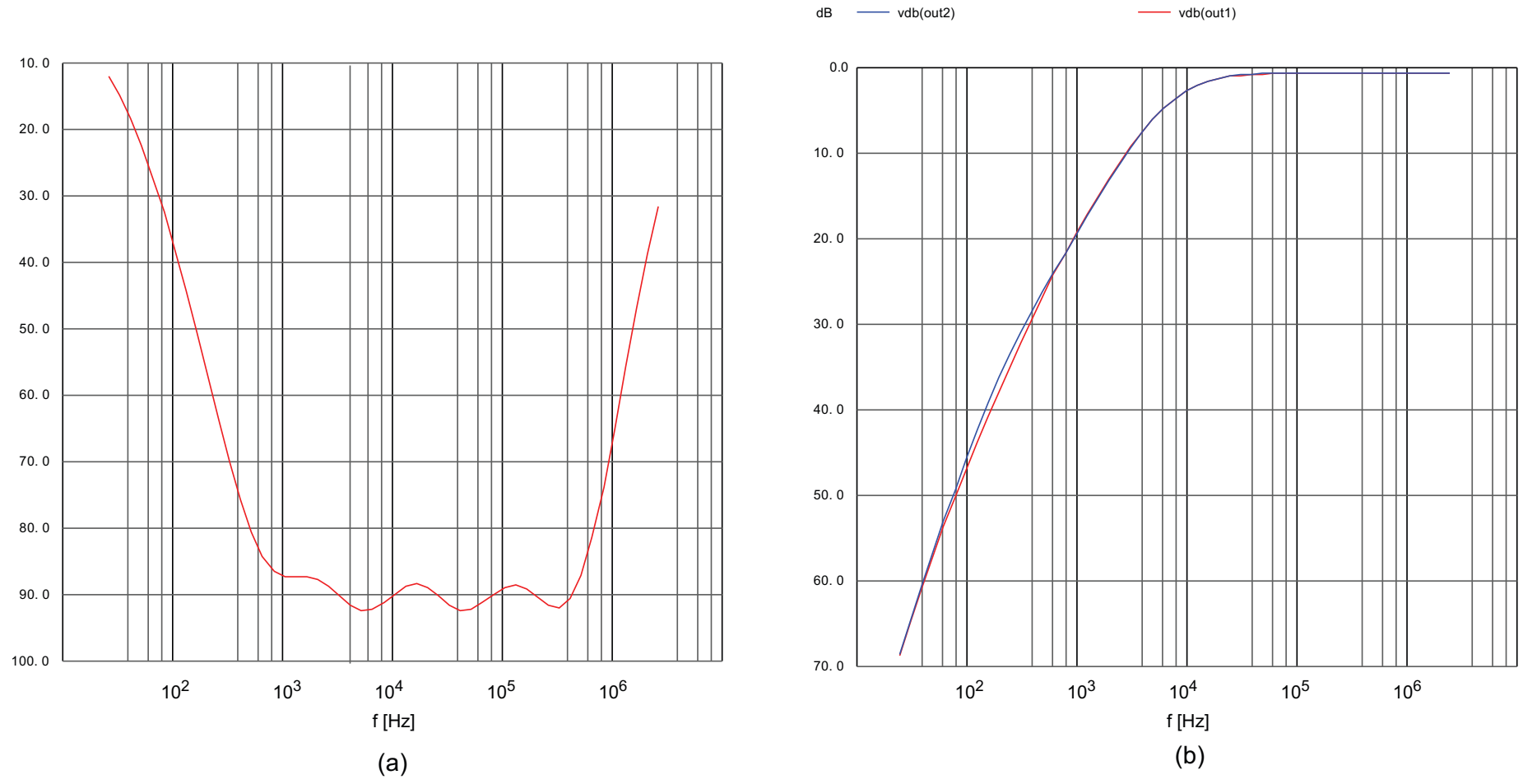

Fig. 6. (a) Difference phase and (b) gain diagrams of the discrete phase shifter network. The phase error within the defined frequency range of 10-300 kHz is below ± 1.5°.

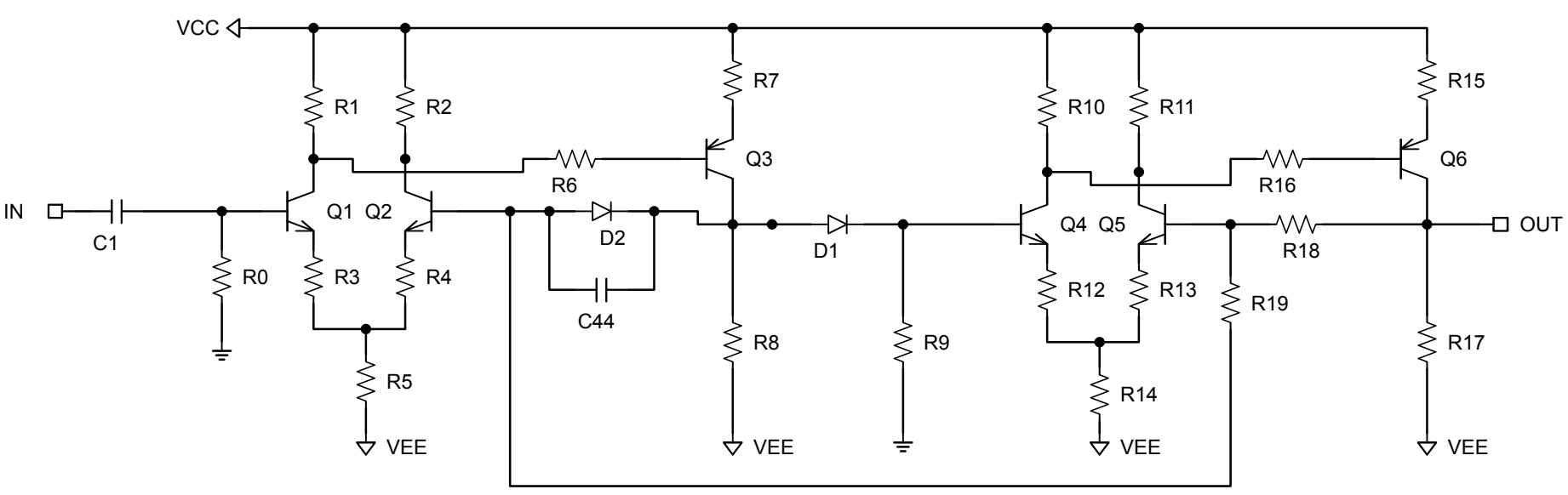

Fig. 7. A full-wave rectifier circuit implementing function $y=abs(x)$.

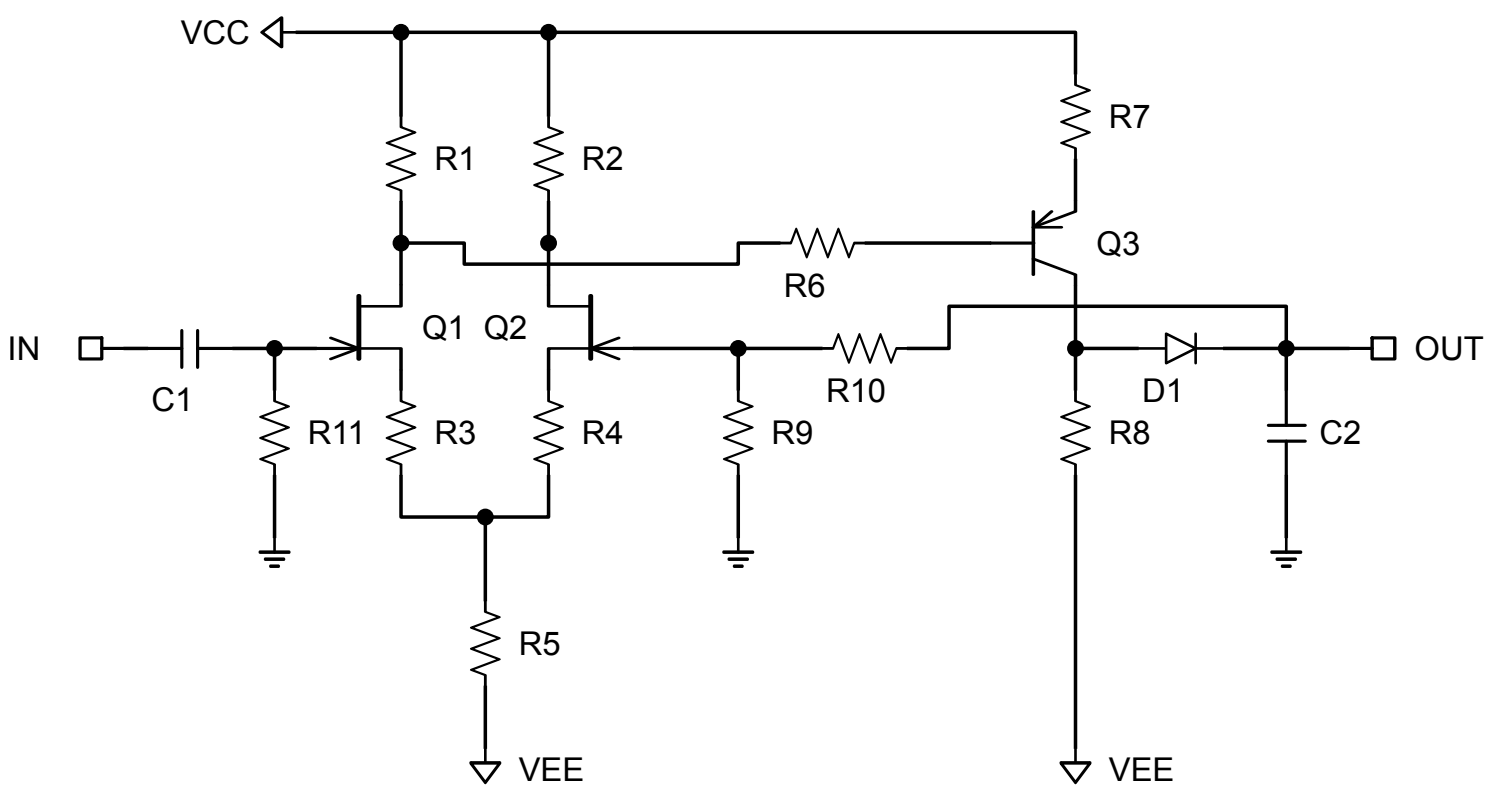

Fig. 8. A peak detector circuit implementing function *y*=*max*(*x*).

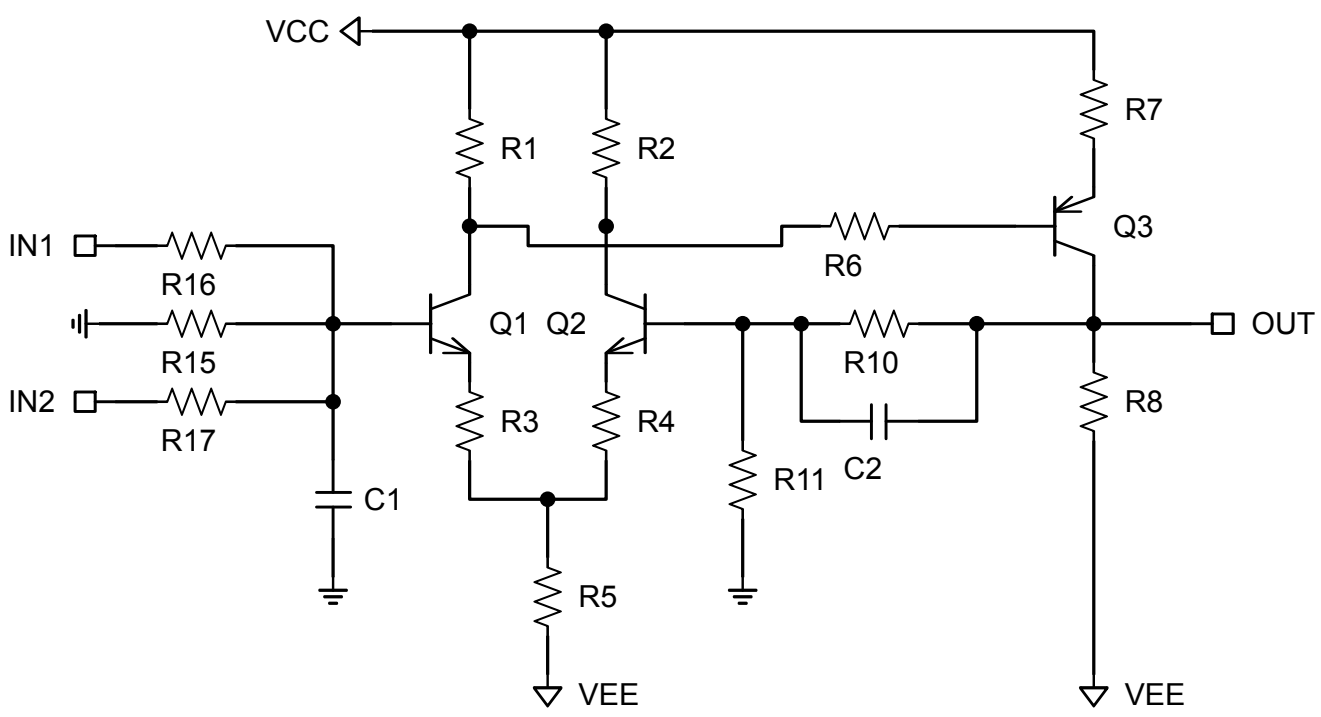

Fig. 9. The mathematical adder circuit implementing the function $y=x_1+x_2$ with additional low-pass (averaging) filtering.

The output of the main part of the analog feature extraction system is shown in Fig. 10. We find a good approximation (error < 10% compared with the mathematical Hilbert transform function) of the signal envelope, with some ripple due to the complex magnitude approximation. The low-pass filters cause a slight phase shift of the envelope signal, which is not critical because all further calculations are bases only on the envelope signal. The peak detector is not fully linear along the full input scale, but sufficient accurate (error below 10%).

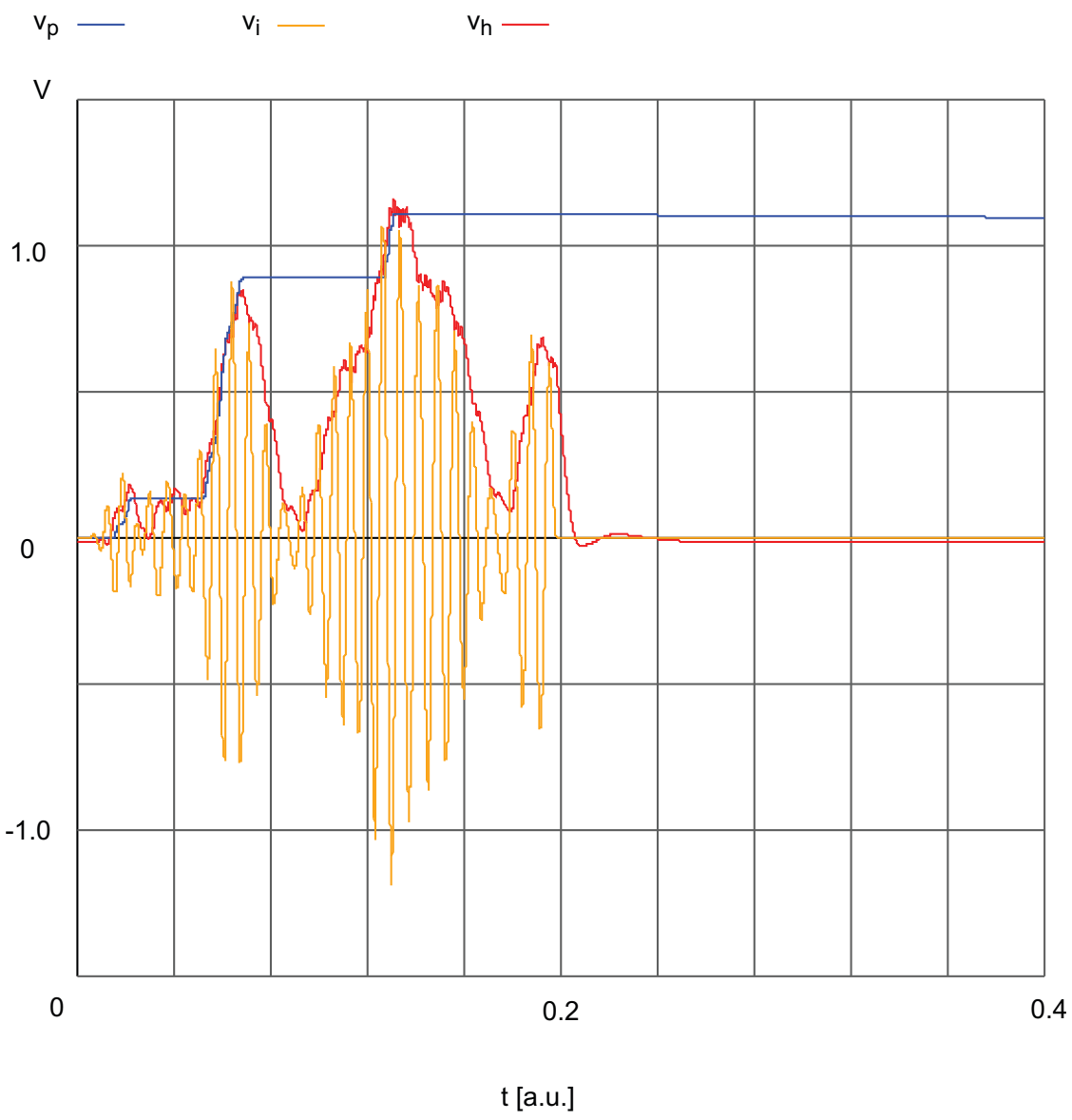

Fig. 10. Output of the entire analog processing system. Input is a Ultrasonic sensor signal $v_{\mathrm{i}}$, output are the approximates envelope signal $v_{\mathrm{h}}$ and the output of the peak detector $v_{\mathrm{p}}$.

## 5. Analog Artificial Neural Network: Combined Classification and Regression

The previously extracted signal features should be used to predict:

1. There is a damage within the scan path, and
2. The distance of the damage from the sending transducer (or the position, more generally).

The first class is a classificator model, whereas the second class is a regression model. Both can be combined in one model requiring a special output encoding, as already shown in early work for a digital ANN [10]. For example, if the normalized output of the model is in the range [0,1], then the range can be split into a classification separation range [0,0.5) and a regression range [0.5, 1]. If the output is below 0.5, then there is no damage detected. If there is a damage, then the output above 0.5 can estimate the position (regression part).

The implementation of an Analog Artificial Neural Network (AANN) was discussed and evaluated in [11,12], not part of this work. Fig. 11 shows the principle analog electronic architecture of one neuron of an AANN. The weight and bias parameters are mapped on resistors of the input amplifiers (current nodes), with one positive and one negative parameter branch (i.e., an input weight is connected to one of the paths depending on the weight sign).

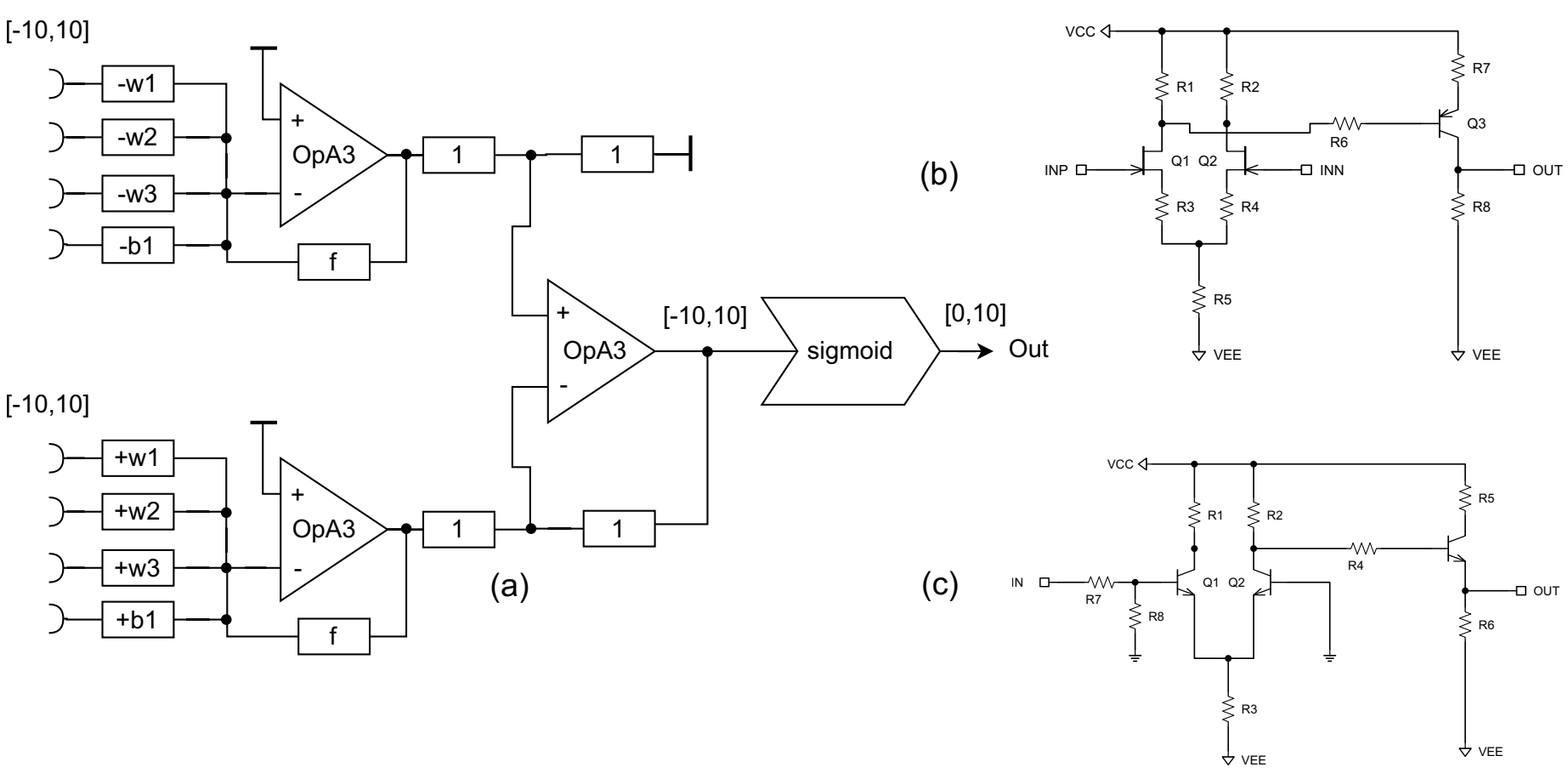

Fig. 11. (a) AANN neuron architecture using three transistor OpAmp circuits OpA3 and the sigmoid function [11,12] (b) OpA3 circuit (c) Sigmoid function circuit (input range [-10V,10V], output range [0,10V]).

## 6. Ultrasonic Damage Detector

To test and evaluate the proposed pure analog damage detector we use simulated sensor signal data. We use a steel plate of 500 by 500 mm, two transducers placed on the top and bottom side of the plate, and a hole of 20 mm diameter as a damage placed along the connecting axis of the transducers, as shown in Fig. 12. It is important to clarify that this is a proof-of-concept experiment using simulated Ultrasonic sensor signals by using the SimNdt2b simulator [13]. The detector is currently specific to this use case and cannot be generalized, and therefore no rigorous statistical analysis is performed.

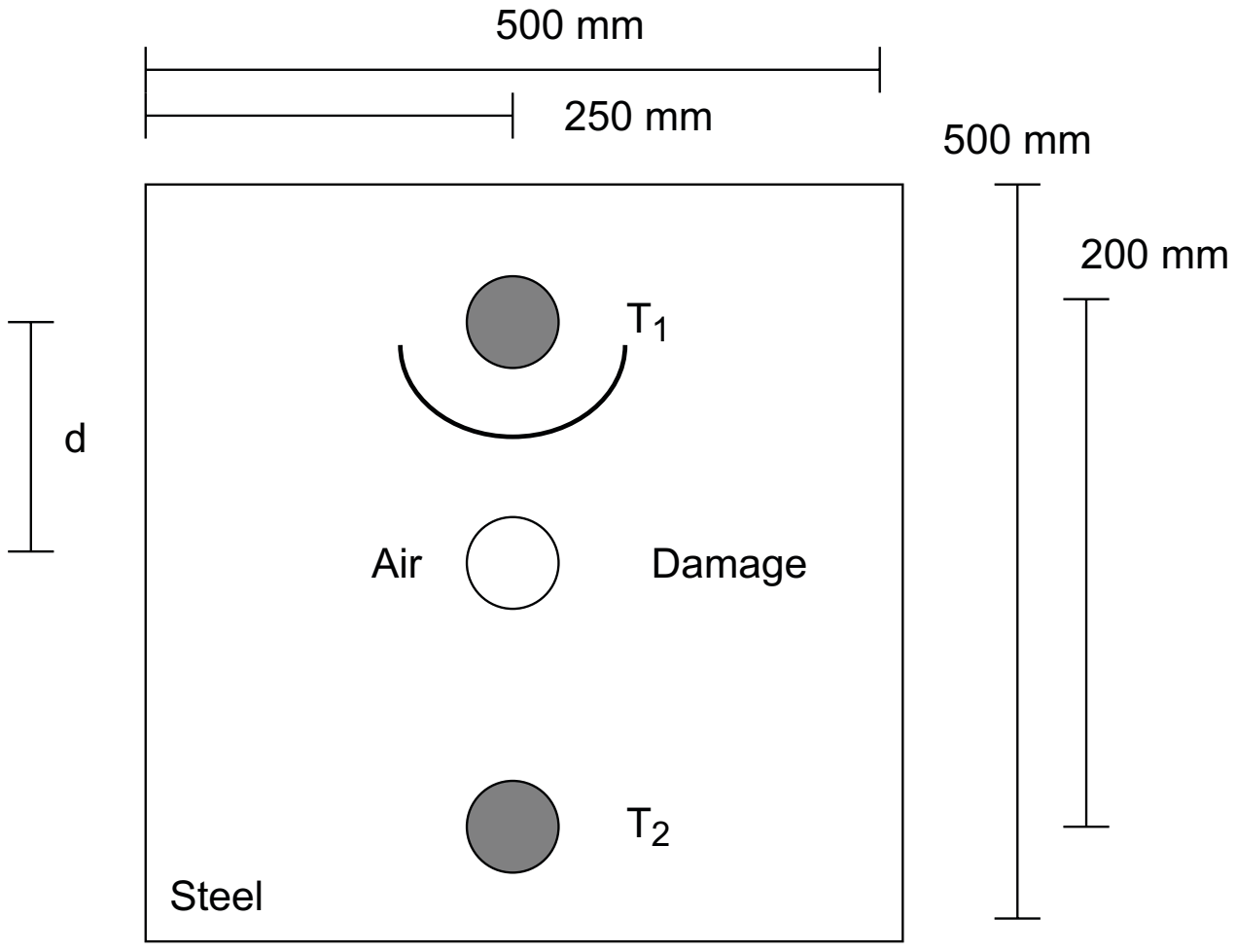

Fig. 12. Experimental demonstrator setup with a steel plate, two transducers PZT and a damage (hole of diameter 20 mm) with varying distance to the sending transducer $PZT_A$.

The pitch signal for the sending transducer is a Gaussian-windowed sine wave with a base frequency of 80 kHz and about 10 cycles to get a sufficient broad main peak.

Fig. 13 shows the response of the analog feature extractor for 20 damage cases and one baseline case without a damage. Shown are the normalized and scaled signal envelope features $T_1$ (time stamp of main peak ), $T_2$ (time stamp at 20% of main peak height), and $A_1$ as the integrated signal area between $T_1$ and $T_2$.

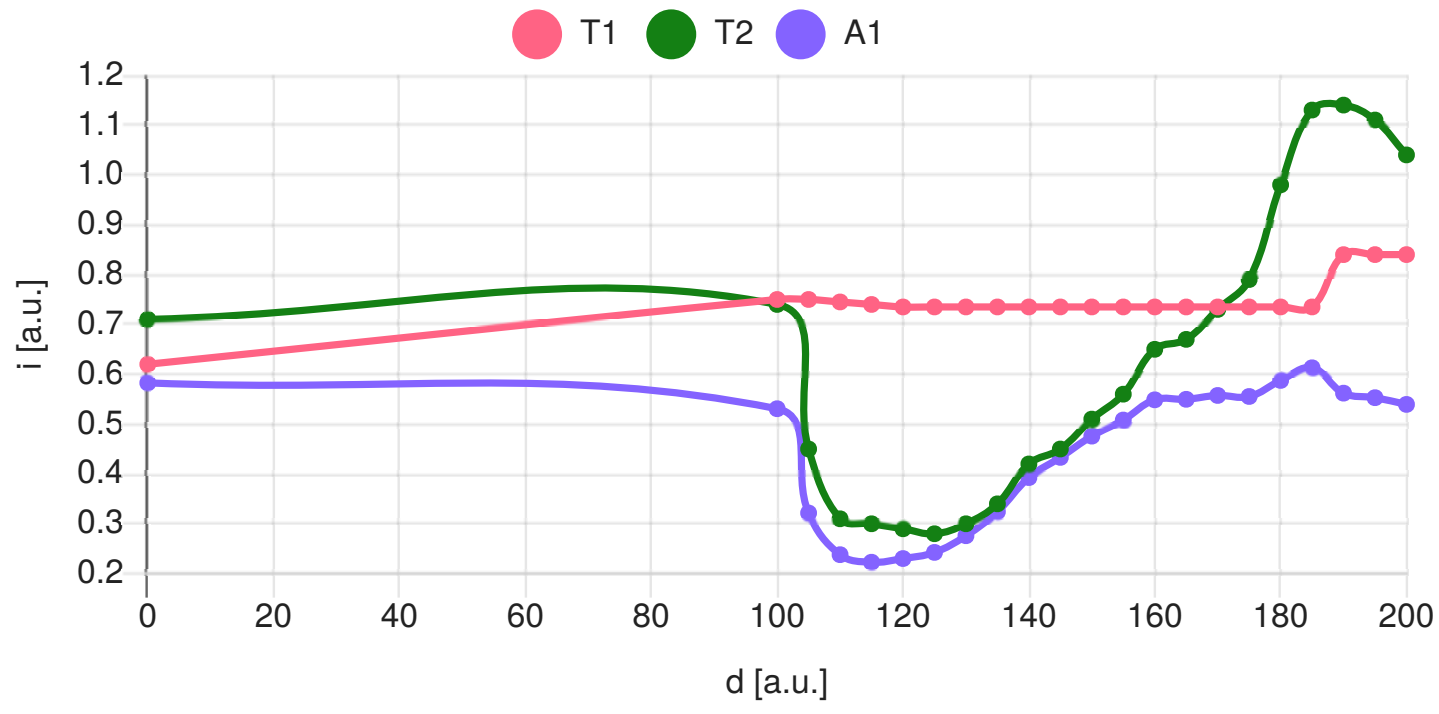

Fig. 13. Normalized and scaled signal features extracted by the analog circuit plotted versa the damage distance (encoded with an offset of 100 mm, d=0 is the non-damage case). The data points are interpolated by splines for visualization only.

The signal features are passed to a simple AANN with only 3 neurons and two layers [2,1]. Each AANN neuron cell use a sigmoid activation function as introduced in Sec. 5. Ex. 1 shows a parameter set as a result of the AANN training (using data-driven numerical methods with the aid of electronic simulation and electronic surrogate models). The weight and bias parameter are basically amplification factors, which must be implemented with the analog electronics. All amplification factors are below 20, suitable for the analog implementation.

Ex. 1. Predictor model parameters for the two layers (Layer 1 is only an input identity layer).

```
Class: mlp
Layers: [3,2,1]
SigmoidLayers: [2,1]
Predictors: 3
=== Bias Layer 2 ===
[1] -10.03  16.44
=== Weights layer 2 ===
        [,1]   [,2]
[1,]  17.62   3.02
[2,]  -9.58  -9.80
[3,]  10.08 -17.96
=== Bias Layer 3 ===
[1] -0.09
=== Weights Layer 3 ===
        [,1]
[1,]  19.85
[2,] -17.85
```

Fig. 14 shows examples of prediction results of the AANN. The output in the normalized range [0,1] is encoded in such way the values in the sub-range [0,0.5) indicate no detected damage, and values in the range [0.5,1] indicate a detected damage and the distance from the sending transducer in the range 100-200 mm.

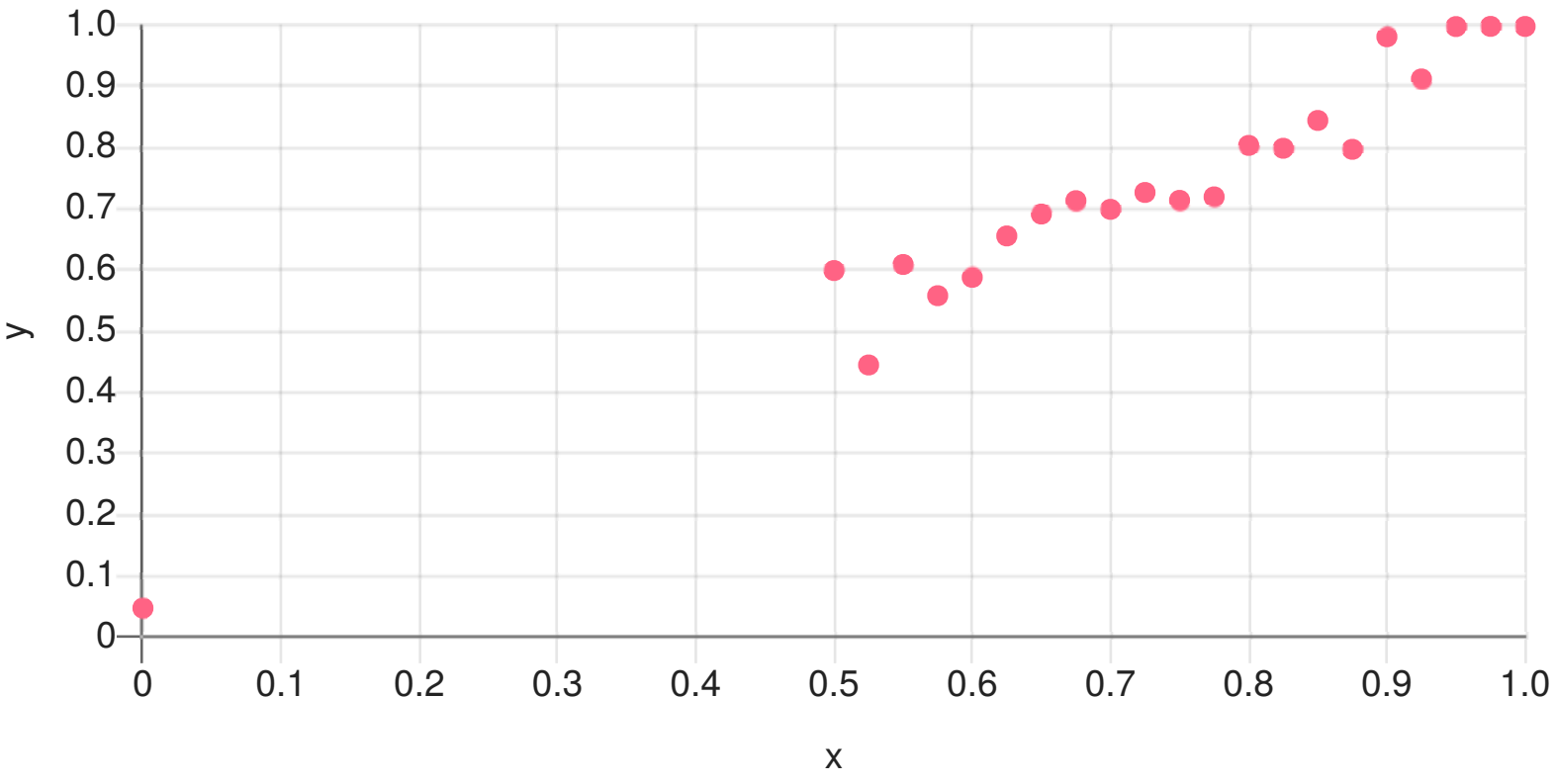

Fig. 14. Some prediction results of the AANN. The x-axis shows the normalized labeled position of the damage (0: no damage, 0.5-1.0 is the detection range for 0-100 mm, respectively), the y-axis shows the predictor output.

## 7. Conclusion

This work discussed methods for extracting features from time-dependent signals, emphasizing the use of the Hilbert transform to compute the analytical signal and its envelope, which is significant in damage diagnostics for Ultrasonic signals. The analytical signal is derived by superimposing the original signal with its complex-valued Hilbert transform, providing a means to calculate the envelope. It details the use of all-pass filters to achieve a 90° phase shift necessary for the Hilbert transform, with a specific design focusing on first-order filters due to practical limitations in accuracy. Key calculations involve determining pole frequencies and estimating phase shift error. An analog signal processing architecture is proposed to extract relevant features of Ultrasonic signals, including peak time, height, width, and area, utilizing controlled analog switches and integrators.

The analog feature extractor primarily utilizes an all-pass filter network, implemented as a fourth-order discrete transistor network. Each filter consists of a single transistor amplifier, with specific resistor-capacitor pairs determining the filter frequencies. The pole frequency is primarily defined by capacitors due to fixed resistor values that establish the amplifier's DC operating point. Second-order filters are preferred over first-order, but interference between multiple RC networks introduces phase errors in a single-transistor setup. The system also employs difference amplifiers and minimal OpAmp circuits for signal feature calculations, demonstrated by designs including an absolute value function and a peak detector circuit. The final output showcases a good approximation of the signal envelope with errors below 10%, albeit with some phase shifts and non-linearity in peak detection.

We demonstrated the functional suitability of a pure analog electronics Ultrasonic damage detector requiring less than 100 bipolar (BJT) and JFET transistors and 500 electronic components. The entire analog processing is able to detect damages within a scan path between two transducers. Moreover, the analog damage detector can estimate the damage distance from the sending transducer with suitable accuracy by using only three signal features. The signal features were derived from the signal envelope using a simple fourth order all-pass filter phase shifter network.

As an outlook the next step is the transition from silicon-based electronics towards organic electronic circuits, more specifically organic electrochemical transistors (OECT), which can be printed on any non-conducting substrate, e.g., by using ink-jet drop-on-demand or screen printing technologies. The entire circuit area can be estimated to less than 400 $mm^2$ (assuming transistor area less than 1 $mm^2$), which outlines the suitability for material-integrated sensing systems (e.g., using system-on-foil technology).

## 8. Appendix A

Electronic component values. All supply voltages are VCC=+10V and VEE=-10V.

| **Component** | **Value** |
|---|---|
| Q11-Q15, Q21-Q25 | 2N3904 |
| R101, R201 | 100 kΩ |
| R102, R292 | 250 kΩ |
| R103, R303 | 1 kΩ |
| R104, R204 | 1 kΩ |
| R11-R14, R21-R24 | 1 kΩ |
| C11 | 14359774 pF |
| C12 | 1200895 pF |
| C13 | 149451 pF |
| C14 | 18148 pF |
| C21 | 3489267 pF |
| C22 | 423722 pF |
| C23 | 52732 pF |
| C24 | 4409 pF |
| R120, R220 | 1 kΩ |
| C10, C20 | 500 nF |

Tab. 2. Component values of the all-pass filter (Fig. 5).

| Component | Value |
|---|---|
| Q1, Q2, Q4, Q5 | 2N3904 |
| Q3, Q6 | 2N3906 |
| R1, R2, R10, R11 | 1 kΩ |
| R3, R12 | 82 Ω |
| R4, R13 | 47 Ω |
| R5, r14 | 1k Ω |
| R6, R16 | 2.2 kΩ |
| R7, R15 | 120 Ω |
| R8, R17 | 680 Ω |
| R9 | 10 kΩ |
| R19 | 0.9 kΩ |
| C44 | 1 nF |
| D1,D2 | DMOD |

Tab. 3. Component values of rectifier circuit (Fig. 7).

| Component | Value |
|---|---|
| J1, J2 | BF245A |
| Q3 | 2n3906 |
| R1, R2 | 1 kΩ |
| R3 | 82 Ω |
| R4 | 0 Ω |
| R5 | 1.6 kΩ |
| R6 | 2.2 kΩ |
| R7 | 120 Ω |
| R8 | 680 Ω |
| R9 | 4000 kΩ |
| R10 | 1000 kΩ |
| R11 | 100 kΩ |
| C1 | 100 nF |
| C2 | 220 nF |
| D1 | DMOD |

Tab. 4. Component values of peak detector circuit (Fig. 8).

| Component | Value |
|---|---|
| Q1, Q2 | 2N3904 |
| Q3 | 2N3906 |
| R1, R2 | 1 kΩ |
| R3 | 82 Ω |
| R4 | 39 Ω |
| R6 | 2.2 kΩ |
| R7 | 180 Ω |
| R8 | 680 Ω |
| R10, R11, R16, R17 | 10 kΩ |
| R15 | 100 kΩ |
| C1, C2 | 10 nF |

Tab. 5. Component values of adder circuit (Fig. 9).

| Component | Value |
|---|---|
| Q1 - Q3 | 2N3904 |
| R1 | 9105 Ω |
| R2 | 9105 Ω |
| R3 | 9407 Ω |
| R4 | 1000 Ω |
| R5 | 908 Ω |
| R6 | 45591 Ω |
| R7 | 51022 Ω |
| R8 | 1328 Ω |

Tab. 6. Component values of sigmoid circuit used in the AANN derived by Particle Swarm Optimization and electronic simulation from the mathematical function (Fig. 11).